# LIINUS/SERPIL: a design study for interferometric imaging spectroscopy at the LBT

F. Müller Sánchez*[a], C. Gál[b], F. Eisenhauer[a], A. Krabbe[b], M. Haug[a], C. Iserlohe[b], T. M. Herbst[c]
[a]Max-Plank Institute for extraterrestrial Physics, Giessenbachstr., P.O.Box 1312, D-85741 Garching;
[b]University of Cologne, I. Physics Institute, Zülpicher Str. 77, D-50937 Köln;
[c]Max-Planck Institute for Astronomy, Königstuhl 17, D-69117 Heidelberg

## ABSTRACT

We present two design concepts and the science drivers of a proposed near-infrared interferometric integral field spectrograph for the LBT. This instrument will expand the capabilities of the currently-under-construction interferometric camera LINC-NIRVANA with spectroscopy by means of an integral field unit (IFU) located inside the LINC cryostat. Two instrument concepts have been studied in detail: a microlens array IFU with a spectrograph built entirely inside LINC (the LIINUS approach), and a lenslet+fibers IFU feeding an external spectrograph (the SERPIL approach). In both cases, the instrument incorporates imaging interferometry with integral field spectroscopy, an ideal combination for detailed studies of astronomical objects down to below 10mas angular resolution in the near-infrared. The scientific applications range from solar system studies and spectroscopy of exoplanets to the dynamics of stars and gas in the central regions of the Milky Way and other nearby galaxies.

## 1    INTRODUCTION

The Large Binocular Telescope (LBT) is an innovative project which by incorporating two 8.4m mirrors on a 14.4m center-to-center common mount offers several interesting possibilities for interferometry. Currently two different beam combiners are being developed for this telescope: the LBTI (Large Binocular Telescope Interferometer [1]), and LINC-NIRVANA (the LBT Interferometric Camera, Near-IR / Visible Adaptive iNterferometer for Astronomy [2]). As LBTI is essentially a nulling interferometer, and LINC can be understood as a high resolution camera, a natural next step for exploiting the full configuration of the LBT would be the construction of an interferometric spectrograph.

Integral field spectroscopy (IFS) has demonstrated to be a very powerful observational technique in modern astronomy. By adding a high resolution integral field spectroscopic capability to the LBT, we will be able not only to exploit efficiently the potential of the telescope, but also to bring several science programs to the next level via interferometry - increased sensitivity (collecting area of an equivalent 12m-telescope), spatial resolution (~16 mas in *K*-band) and number of spectral elements (R>3000).

This manuscript describes developments towards the design of a high-resolution integral field spectrometer performing Fizeau interferometry at near-infrared wavelengths in the LBT. The design study was initiated at the I. Physics Institute of the University of Cologne under the name of LIINUS (LINC-NIRVANA Interferometric Imaging Near-infrared Upgrade Spectrograph). A similar project was independently launched at the Max-Planck Institute for extraterrestrial Physics (MPE) in Garching under the name of SERPIL (Spectrograph for Enhanced Resolution Performing Imaging interferometry on the LBT). The joint cooperation between the two groups is reflected in the composite acronym of the project: LIINUS/SERPIL or also referred as SERPIL/LIINUS.

This document is organized as follows: Section 2 describes the wide range of science that will be tackled with this instrument focusing on two key science cases: nearby AGN and the Galactic Center. In Section 3 we show the top-level instrument requirements that follow from the key science drivers. The two design concepts studied in detail are presented in Section 4. Finally, the conclusions of the investigations are summarized in Section 5.

*frankmueller@mpe.mpg.de; phone +49 089 30000-3587; fax +49 089 30000-3569; http://www.mpe.mpg.de

# 2   SCIENTIFIC MOTIVATION

This section describes the scientific rationale for building LIINUS/SERPIL. An introduction of the exciting possibilities that this instrument will offer has been discussed in [3]. The science cases range from binary stars, exoplanets, star forming regions and star clusters, to intermediate mass black holes (IMBH), the Galactic Center, nearby galactic nuclei and in particular nearby AGN. Perhaps the most important foreseeable application of LIINUS/SERPIL will be studies of stars and gas in the immediate vicinity of supermassive black holes (SMBH) in nearby AGN, with the ultimate goal of testing the Unified Model of AGN [4]. This key science case drives the instrumental design. Therefore, we will present in this document an in-depth study of the AGN science cases. In addition, we will proceed in somewhat less detail with our second key science case, the Galactic Center.

## 2.1   Nearby Active Galactic Nuclei

Active Galactic Nuclei (AGN) are the most luminous sources of electromagnetic radiation in the universe, and as such they represent one of the most interesting areas of research in astronomy. Over the last few years, thanks to adaptive optics (AO) observations with 8-10 m telescopes, it has become possible to begin probing the nuclei of nearby active galaxies below 100 mas scales. Such objects, at distances of a few to a few tens of Mpc, are almost exclusively Seyfert galaxies, which are less luminous then QSOs but are believed to fit into the same unification scheme. AO observations of these sources have revealed many new insights, in particular the prevalence of star formation in the central few tens of parsecs and the way in which it may influence the fuelling of the BH, as well as the distribution and kinematics of the molecular gas and how it relates to the torus.

These crucial topics in the field of AGN research could be better understood by performing integral field spectroscopic observations at the interferometric focus of the LBT with LIINUS/SERPIL. In the closest AGN – those within 20 Mpc – one will be able to probe scales less than 1.5 pc, a scale which should resolve the putative torus and will bring us close to the Broad Line Region of the nearest objects.

### 2.1.1   Nuclear star formation

The spectroscopic capability of LIINUS/SERPIL provides an excellent opportunity to extend the work done on AGN with current integral field spectrographs to much smaller scales (e.g. [5, 6, 7, 8]). One particular aspect that would have important consequences is the star formation. Recently, Davies et al. [5] used SINFONI [9] to show that in a number of AGN the stellar luminosity increases towards the nucleus; and that these nuclear starbursts, while still young (10-300 Myr), are no longer active. Comparing the distribution and kinematics of the stellar continuum with that of the molecular gas it seems likely that the star formation actually occurred within part of the torus. Given that the star formation occurs on scales of <50 pc, it is inevitable that it and the AGN will have some mutual influence on each other.

There exist increasing observational and theoretical evidences that the AGN and the surrounding star formation are inextricably associated and probably interact which each other on different scales. On the observational side, Davies et al. [5] concluded that the starburst (on very small scales) has a considerable impact on the AGN fuelling. Their results, summarized in Fig. 1, hint to a possible relationship between the characteristic age of the star formation and the accretion rate onto the AGN. The AGN which are radiating at lower efficiency $<0.1$ $L/L_{Edd}$ are associated with starbursts younger than 50-100 Myr; AGN that are accreting and radiating more efficiently $>0.1$ $L/L_{Edd}$ have starbursts older than 50-100 Myr. This implies that there could be a delay between starburst activity and AGN activity. They have interpreted this as indicating that fuelling a supermassive black hole requires the presence of a starburst.

On the theoretical side, hydrodynamical simulations of the impact of stellar evolution on gas in the central 50 pc of AGN have led Schartmann [10] to a remarkably similar conclusion to that above. He found that supernovae can blow low density gas away, leaving long dense filaments (Fig. 2). The interplay between these filaments and the slower stellar ejecta lead to inward accretion to form a central turbulent disk. This clearly indicates that a starburst does indeed have a dramatic impact on the nuclear region, and that gas ejected in slow winds can be accreted along filaments.

Making use of the CO2-0 bandheads at 2.3μm in *K*-band spectra from LIINUS/SERPIL would extend this work to scales a factor of ~4 smaller. At first, we will be able to resolve the central stellar cluster on scales of a few parsec from the nucleus, and probably in some objects, separate individual star-forming regions. In the nearest objects (5-10 Mpc), SERPIL/LIINUS will enable one to test if there still exists the signature of stars in the central 0.5-5 pc. If so, spectral synthesis models can then be used to infer the age and properties of the stars in this region, which corresponds to that of the putative torus. As the starburst would be occurring within the torus, one will have to tackle the following question: is it possible for stars to form in such a turbulent environment? At last, using LIINUS/SERPIL in a large sample will enable one to address the important issue of the AGN-starburst connection.

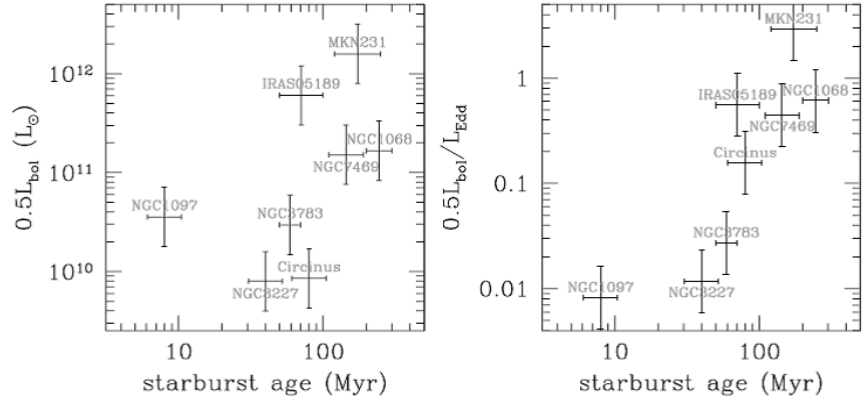

Fig. 1. Relationship between the luminosity of an AGN and the age of the most recent episode of nuclear star formation [5]. In the left panel the luminosity is shown in units of solar luminosities. In the right panel the luminosity is shown with respect to the Eddington luminosity of the black hole.

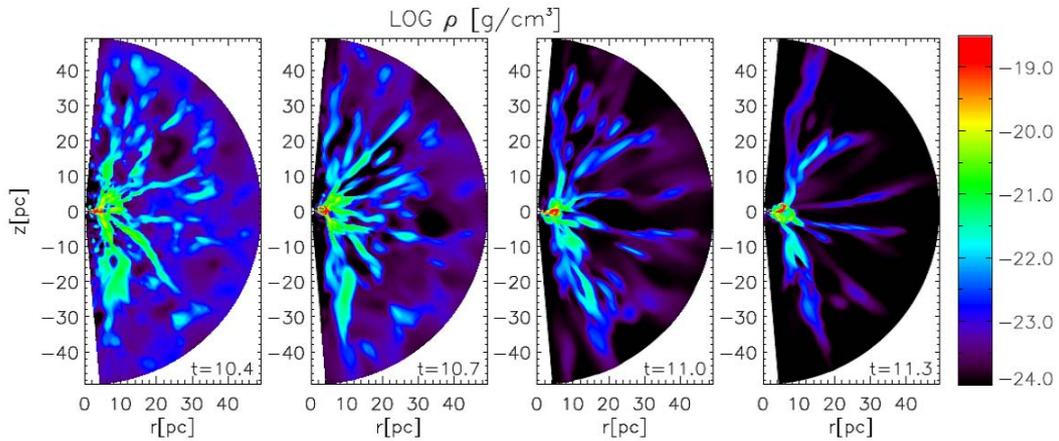

Fig. 2. Hydrodynamical simulations of the impact of stellar evolution on gas in the central 50 pc of AGN [10].

### 2.1.2 The obscuring molecular/dusty torus

A fundamental prediction of the unified model of AGN is the existence of dense molecular gas and dust (T~100-1500 K) in the form of a geometrically thick (z >> r) and optically thick torus responsible for the aspect-angle and wavelength-dependent UV (and perhaps X-ray) obscuration of the nucleus. So far, most of its properties are poorly understood. As previous attempts to obtain a clear image of it have failed, its characteristics have been inferred only indirectly, leading to a controversial debate about its extent and physical properties.

Recent mid-IR VLTI results with MIDI have begun to shed light on the physical characteristics of this structure. Currently 10 AGN have been observed with MIDI, and the brightest 2 – Circinus and NGC 1068 – have yielded detailed results [11][12]. In both cases the most compact component is highly elongated with a major axis length of ~1 pc. Around these are larger more symmetric components on scales of a few parsecs. These results already imply that, in order to have such warm dust so close to the AGN, the torus must be clumpy. The observations agree remarkably well with clumpy models of compact torii [13]. In addition, the hydrodynamical simulations of [10] (Fig. 2) show that the torus may actually have substructure: a larger scale diffuse part consisting of long filaments, and a small scale dense turbulent disk. It would be the former structure that the AO data from 8-10 m telescopes have found; and the latter that shows up most prominently in interferometric observations of AGN. This is demonstrated in Fig. 3 for the particular and very important case of NGC1068. In the left panel, SINFONI observations of the 2.122μm H$_2$ 1-0S(1) emission in the central 0.8"x0.8" of NGC1068 with a resolution of 0.075" (~5 pc) is presented [7]. This line probes hot (~1000 K) and moderately dense (~10$^3$ cm$^{-3}$) molecular gas and as such, may be an excellent tracer of gas in the nuclear region. Its spatial distribution can not only expose the potential presence of the torus, but also provide physical information on fueling or feedback mechanisms. Müller Sánchez et al. [7] interpreted their integral-field data at these scales as tidally disrupted streamers that forms the optically thick outerpart of an amorphous clumpy molecular/dusty structure which contributes to the nuclear obscuration. On the right panel, an overlay of the new interferometric mid-IR observations of NGC1068 [12], the H$_2$O maser disk and the nuclear radio continuum at 5 GHz is shown. Jaffe et al. [12] found that the fitted Gaussian components to the (u,v) plane of the central 10μm source resemble a disk similar to the H$_2$O masers.

This example demonstrates that adaptive optics observations of nearby AGN have started to give new insights to the global structure of the torus. Observations with LIINUS/SERPIL will show how these large scale properties are reconciled with the small scale structures inferred from interferometric observations. By combining analysis of the warm molecular gas traced by the H$_2$ 1-0S(1) ro-vibrational emission with near-IR continuum measurements at scales of 0.5-5 pc from the nucleus, one can perform much more detailed studies and use the data to constrain the torus model parameters (inner and outer radius, inclination, thickness, clumpiness, etc).

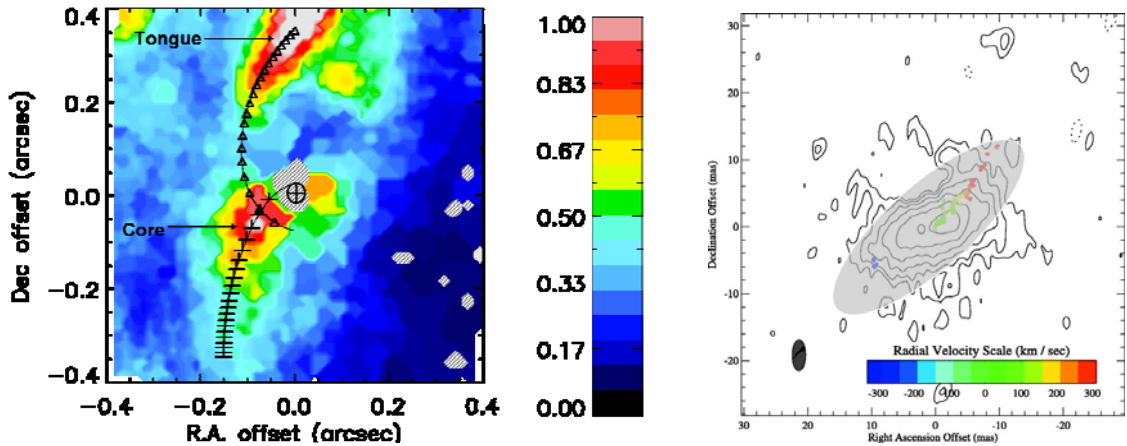

Fig. 3. Left: Flux map of the H$_2$ emission in the central 0.8"x0.8" of NGC1068 [7]. The open triangles show the projected trajectory of the northern streamer of gas. The half-crosses show the past trajectory of the gas which is currently located in front of the AGN. Right: Components of NGC 1068 superimposed on VLBI radio data [12]. The contours show emission at 5 GHz, the grey ellipse the position of water maser emission and the 10 μm source. Notice that the scale of the image in the right panel is ~10 times smaller compared to the one in the left. SERPIL/LIINUS with a resolution of 16 mas will be able to resolve the structures in the right image and connect them with those at larger scales.

### 2.1.3 Black hole masses

The third issue motivating an interferometric integral field spectrograph for the LBT is the determination of the mass of the central SMBH from resolved nuclear stellar dynamics in AGN. The masses of black holes in nearby AGN are most commonly estimated by reverberation mapping [14]. Providing an independent measure of $M_{BH}$ for reverberation masses would allow one to begin to understand the geometry of the broad line region. In addition, estimates of $M_{BH}$ are important to acquire a more precise knowledge of the behaviour of the correlation between $M_{BH}$ and the bulge velocity dispersion (the $M_{BH}$-σ relation [15]). It is generally accepted that the $M_{BH}$-σ relation should be valid for all spheroids irrespective of their nature (spiral, bulge, pseudo-bulge, elliptical or irregular) and the nature of the black hole (quiescent, active, supermassive, intermediate mass, stellar). However, almost without exception the $M_{BH}$ estimated from stellar kinematics have been derived only for nearby bulge dominated E/S0 galaxies. The smaller bulges of spiral Seyfert galaxies imply lower $M_{BH}$, making it difficult to spatially resolve the stellar kinematics. In addition in these galaxies the glare of the AGN itself overshines the stellar luminosity adding another obstacle to the determination of the $M_{BH}$.

Integral field spectroscopy is extremely well suited for studying the inner dynamics of the stars in nearby AGN and therefore the determination of the nuclear mass, since it helps to avoid ambiguous interpretations as from long-slit spectroscopy. The reason, explained clearly by [16], is that long-slit data may miss some orbits if the (localized) regions of high projected surface brightness do not happen to fall within the slit. As a result, a model derived using long-slit data may be significantly different from that found from integral field data. The high spatial resolution, sensitivity and integral-field capability of LIINUS/SERPIL will provide an ideal combination to do this. With its 16 mas angular resolution in *K*-band, LIINUS/SERPIL will be able to resolve the radius of influence of the black hole in a typical Seyfert galaxy harbouring a $M_{BH} \sim 10^7\ M_{sun}$ out to a distance of ~50Mpc.

### 2.1.4 Expanded samples

By tracking on the AGN itself, one will be able to observe a number of important and well studied objects that have also been successfully targeted with adaptive optics. An initial nearby AGN sample is listed in Table 1. Source selection was driven by technical considerations, the primary criteria being:

1. the nucleus should be bright enough for fringe-tracking (*K*-band magnitude <10 within 1" aperture),
2. the galaxy should be easily observable at LBT's latitude,
3. the galaxy should be close enough so that small nuclear structures can be resolved at the near-infrared diffraction limit of an equivalent 28m-telescope, and
4. the galaxies should be "well-known" so that complementary data can be found in the literature

Table 1. List of 10 sources suitable for observations with LIINUS/SERPIL.

| Galaxy | RA J2000 | Dec J2000 | Nucleus | Dist (Mpc) | K (mag) |
|---|---|---|---|---|---|
| NGC1068 | 02:42:40.7 | -00:00:48 | Sy2 | 14.4 | 8 |
| NGC2841 | 09:22:02.6 | +50:58:35 | L/Sy1 | 9.8 | 8.6 |
| NGC3627 | 11:20:15.0 | +12:59:30 | Sy2 | 8.9 | 8.5 |
| NGC4151 | 12:10:32.6 | +39:24:21 | Sy1.5 | 14 | 9.15 |
| NGC4258 | 12:18:57.5 | +47:18:14 | L/Sy1.9 | 6.95 | 9.74 |
| NGC4579 | 12:37:43.6 | +11:49:05 | Sy2 | 20 | 8.6 |
| NGC4594 | 12:39:59.4 | -11:37:23 | L/Sy2 | 13.7 | 8.52 |
| NGC4725 | 12:50:26.6 | +25:30:03 | Sy2 | 17.1 | 9 |
| NGC4826 | 12:56:43.7 | +21:40:52 | Sy2 | 5.6 | 8.34 |
| NGC5033 | 13:13:27.5 | +36:35:38 | Sy2 | 13.3 | 9.1 |

These 10 objects in themselves will already yield invaluable insights into the detailed structure of AGN. However, it is always important to build statistically significant samples. The fringe tracking capability of LINC/NIRVANA can also observe AGN with nearby reference stars. A cross-correlation of the Sloan Digitised Sky Survey with the ROSAT All-Sky Survey has already been performed, generating a catalogue of X-ray bright AGN suitable for AO [17]. The search yielded a sample of 78 galaxies at redshifts < 1 with compact hard X-ray emission, which have a bright reference star within 40". As such it represents a first step towards assembling a statistically significant sample of galaxies which can be systematically addressed using LIINUS/SERPIL.

## 2.2 The Galactic Centre

Near-IR observations of the stellar orbits in the Galactic Centre have provided convincing evidence for the presence of a supermassive black hole at the position of SgrA* [18]. Radial velocity measurements put further constrains on modeling the orbits of S-stars and hence, allow a more accurate determination of the black hole mass. High angular resolution integral field spectroscopy achieved with LIINUS/SERPIL will help to measure absorption lines (CO bandheads and HeI at 2.17μm) with high contrast. Precise classification of the stars according to their mass by means of spectroscopic measurements will result in a better description of the stellar dynamics of the central star cluster in the immediate vicinity of SgrA*. Such a spectroscopic classification was published by Martins et al. [19] using the S2 B-star as illustrated in Fig. 4.

Furthermore, the following interesting science topics could also be addressed with LIINUS/SERPIL:

- "Paradox of youth", origin and evolution of the surprising number of young, massive stars near to the SMBH,
- Interaction of high velocity winds from young massive stars with the interstellar medium (bow shocks),
- Investigation of counter-rotating disks of young stars,

In conclusion, LIINUS/SERPIL measurements will give unique insights into the three dimensional structure, dynamics and evolution of the stars and gas in the immediate vicinity of the SMBH in the Galactic Centre.

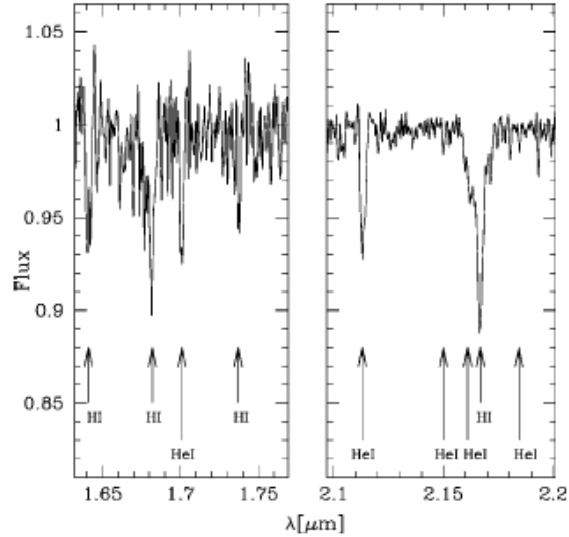

Fig. 4. Spectrum of S2 after summation over all SINFONI observations since 2004. The main absorption lines are indicated.

## 3 TOP LEVEL REQUIREMENTS

The following top-level requirements arise mainly from the science drivers discussed in Section 2. They should provide a realistic picture of the most important parameters of the instrument and serve as a starting point for more detailed trade-off studies and for the development of a technology roadmap.

- Instrument type: near-IR integral field spectrometer in Fizeau interferometric imaging mode
- Field of view: 1"x 1" in interferometric mode
- Science wavelength range: 0.9 - 2.4 μm with optimization to *K*-band
- Spatial sampling: ≤30 x ≤10 mas (2-3 pixels per central fringe)
- Spectral resolution: R > 3000 for OH lines suppression
- Pixel size: 15 μm
- Detector array: 4096 x 4096 pixels
- Throughput > 25%

In addition, working at near-infrared wavelengths requires the spectrograph to be cooled down to about 77 K. However, depending on the instrument configuration (see Section 4), part of the instrument might work at ambient temperatures.

## 4 DESIGN CONCEPTS

The several options for IFS in the context of LIINUS/SERPIL are discussed in a previous publication [3]. Soon after the project was started, it became obvious, that LIINUS/SERPIL would not replace LINC at the telescope and therefore it will be built within the LINC/NIRVANA hardware. In this manner LIINUS/SERPIL will be able to use existing subsystems of LINC to perform important and common tasks such as wavefront sensing and correction, and fringe & flexure tracking. However, strong size constrains are imposed, making the design of LIINUS/SERPIL very challenging.

This focused our investigations to the technical solutions based on an IFU located inside the LINC cryostat. The spectrograph could then be placed either inside LINC or in a new cryostat by transporting the light via optical fibers. Since image slicer IFUs normally occupy a lot of space due to the long pseudo-slit formed [3], for a spectrograph fully integrated to LINC, a lenslet array IFU is the baseline. If the spectrograph is built in a separate cryostat, a lenslet+fibers IFU is the only realistic option. Based on all this, two cases were considered for further studies: the lenslet+fibers approach (or SERPIL approach) with an external spectrograph, and the whole integral field spectrograph inside LINC approach (or LIINUS approach) based on a microlens array IFU.

The two design concepts of LIINUS/SERPIL have different potential show-stoppers: the physical problem of coupling light from a double pupil to optical fibers in the case of the lenslet+fibers approach, and the design of a very compact integral-field spectrograph for the "inside LINC spectrograph" approach. The main results of our investigations on the coupling of light to optical fibers will be discussed in Section 4.3. Regarding to the "inside LINC spectrograph" approach, our investigations suggest that there is space available for upgrades in the upper part of LINC. This space was initially considered as a cylinder of 30 cm diameter, and length of 70 cm. In order to demonstrate that building a spectrograph inside this volume is feasible, a preliminary optical design employing a microlens array IFU was first developed [3]. However, as this volume has changed over the last two years, a new design of the spectrograph has been developed. In addition, as the LINC's focal plane has to be re-imaged at the entrance of the spectrograph section, an optical path inside LINC which brings the light from LINC's original focal plane at the bottom of the cryostat to the upper part has been identified. A preliminary design of this and the spectrograph will be presented in Section 4.2.

### 4.1 Common modifications to LINC for SERPIL/LIINUS

Some modifications have to be done to LINC/NIRVANA in order to shape up the necessary mechanical and optical interfaces for LIINUS/SERPIL. One of these changes is in the filter wheel, where at one of the filter positions an additional folding mirror is integrated (see Fig. 5). The small mirror picks up the 1 x 1 (or 2 x 2) arcsec beam required for both instrument concepts and reflects toward the input optical interface of the IFUs. The careful design of the pick-up mirror considers the baffling structure of the filter-wheel, the available space in the cryostat, and the requirement that the relayed focal plane can be easily reached through the service hole of the LINC cryostat. These requirements consequently define the tilt angle of the pick-up mirror, which is 54.75° normal to the LINC science beam. Since only 10 or 20 % of the LINC beam is used for the IFU, the rest of the beam passing through a given filter and reaching the LINC detector can be used for calibration purposes. For the LIINUS/SERPIL beam a hole has to be cut in the baffling structure. During LINC operation this hole remains covered and will be opened for observations with the IFU.

The fibre head of SERPIL or the pre-optics of LIINUS are mounted on a service plate that is fixed to the mounting structure of LINC. The rigid design performs a very stable platform, which is needed for both of the IFU approaches. The available volume within the LINC cryostat is modelled and the intended allocated volume can be seen in Fig. 6.

### 4.2 The complete spectrograph inside LINC approach (the LIINUS design concept)

The LIINUS concept corresponds to the "inside LINC" approach. The design is driven by scientific, technical, and space requirements. This approach is based on the "TIGER" type of IFUs [3, 20] with an additional stage performing anamorphic magnification.

#### 4.2.1 Re-imaging optics

The interferometric PSF of the LBT, having different dimensions in the high and low resolution directions, could be imaged with circular symmetry onto the square pixels of the detector (see Fig. 10). Thus, the telescope image must be anamorphically magnified in order to exploit the maximal capability of the detector.

Different types of anamorphic magnifying optical systems were studied including prism pair anamorphs and anamorphic refractive optical lens systems [21]. For LIINUS a pure cylindrical lens design is preferred having purely spherical surfaces of each lenses. The anamorphic lens system adopts the focal length of the telescope so that at the focal plane of the anamorphic optics the FWHM dimension of the beam matches the pitch of the field sampling device at about 5x5mas. In the meridional plane (high resolution) this requires an effective focal length of 110770 mm, corresponding to a magnification of 10 over the LINC focal plane. In the sagittal plane (low resolution) the effective focal length results in 30129mm with the associated magnification of 2.72 of the LINC image. The beam speed at its focus is f/320 and f/87 in the high and low resolution directions, respectively.

The anamorphic optical lens system has already been optimized for best image quality, proper magnification, necessary optical path, exit pupil plane transformation into infinity, and least numbers of lenses. The folded optical system consists of 6 cylindrical lenses: 3 are active in the sagittal-, and 3 in the meridional plane, respectively. The folded design fits the allocated volume in the cryostat and also gives place for the image rotator, the K-mirror (see Fig. 7). The entire pre-optical system is accommodated in a folded pipe system, which is fixed to the mounting structure of LINC (Fig. 6).

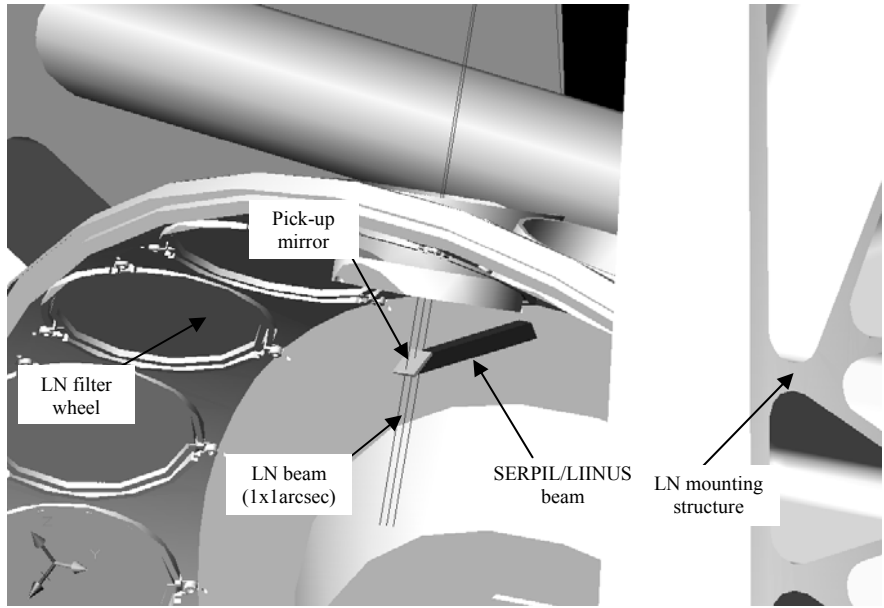

Fig. 5. Position of the pick-up mirror for the SERPIL/LIINUS IFU.

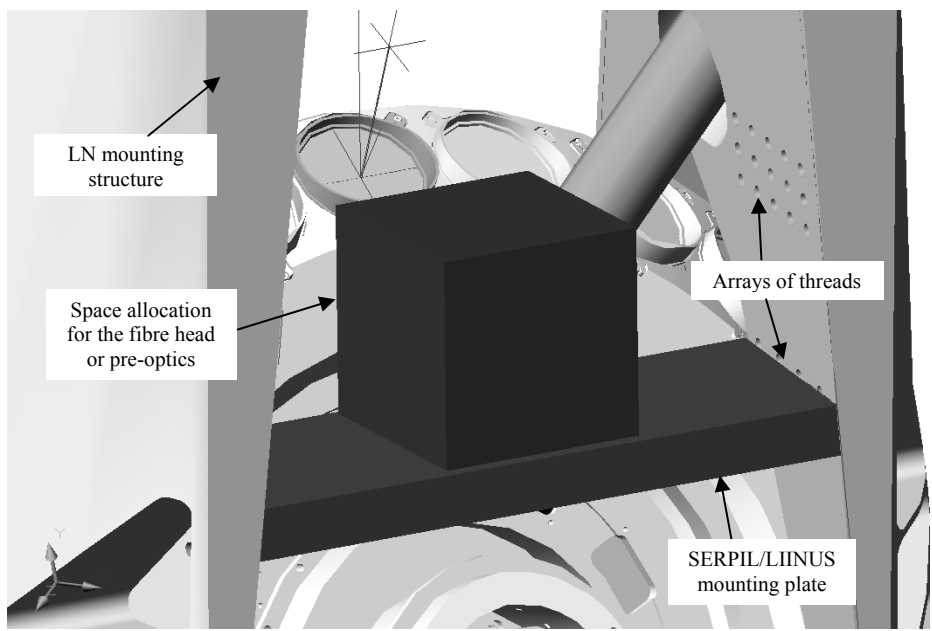

Fig. 6. Mounting plate for the fibre head of SERPIL or the pre-optics of LIINUS. Arrays of threads in the LINC mounting structure are foreseen for height adjustments and for additional fixing points for further parts of the spectrograph.

### 4.2.2 The Integral Field Unit

Microlens arrays are good alternatives for IFUs since their sizes are small and provide good covering efficiencies [3]. A 16 x 64 square shaped microlens array with 180μm pitch is chosen as IFU. The 180μm size corresponds to the FWHM of the PSF at the focal plane so that the lens size guarantees Nyquist-sampling at the design wavelength. The material of the microlenses is fused silica due to its adequate manufacturing properties.

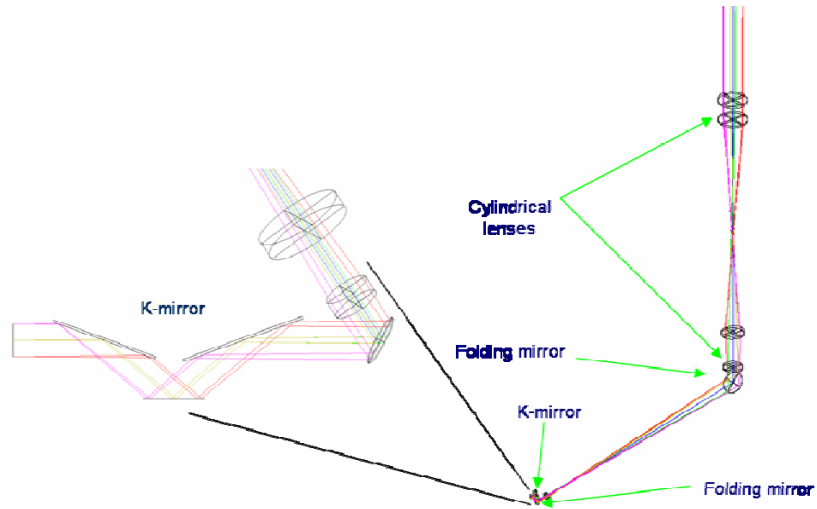

Fig. 7. CAD model of the anamorphic optics design. The optical system comprises the cylindrical lenses, two folding mirrors, and a K-mirror. Due to the small field of LIINUS, the K-mirror might be placed near to the relayed LINC-NIRVANA focal plane, where the deformation tolerances are the smallest.

### 4.2.3 Spectrograph Section

The entire spectrograph including all of its components has a total optical track of about 485 mm and the maximal lens diameter is 40 mm. The instrument with these dimensions fits in the upper part of the LINC cryostat. The optical system has to be folded at least once. A possible arrangement of the spectrograph is presented in Fig 8.

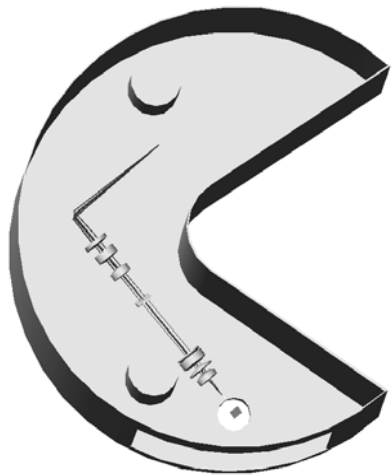

Fig. 8. The space available inside the LINC cryostat as seen from the top. The crescent-moon (or Pac-man) shape offers the possibility to fold the light to fit a spectrograph with a total optical track of ~1.0 m.

**Collimator:** Collimating optics are required to properly illuminate the dispersive element of the spectrograph. The necessary focal length is 140mm in order to achieve the required 3500 resolving power. Good optical performance was attained with 4 lenses made of SF56A and $BaF_2$. Both of these materials are available and do not introduce manufacturing difficulties.

**Dispersive element:** The spectral resolution of LIINUS can be achieved by using a grism - a prism with a grating on one of its surfaces. Grisms have the advantage that by using higher diffraction orders (m = 3, 4, 5), the bandwidth of the instrument will cover the entire J, H, and K bands. The grooves profile is prefered to be sawtooth in order to take advantage of the maximum grating efficiency.

**Camera and Detector:** Based on the requirements of the spectrograph a camera objective was designed. A 3-lens design promises good optical efficiency and guarantees proper imaging quality. The lenses are made of SF56A and $BaF_2$.

LIINUS is designed for a 2k x 2k (or 4k x 4k) 18 (or 15) micron pixel array. Quantum efficiencies are expected to be greater than 80% in the near-IR with a read-out noise of about 2 electrons. The spectrum width is about 2 pixels on the detector. A rotation of the detector is needed for an optimum packaging of the spectra.

**4.3 The lenslet + fibers approach (the SERPIL design concept)**

The principle of the lenslet+fibers IFS for LIINUS/SERPIL is shown schematically in Fig. 9. The telescope focal plane is imaged on the front curved surface of the microlenses. Each microlens forms a telescope pupil image on its back flat surface, where the fiber is attached. Each micropupil is then transported by fibers from the LINC hardware to another cryostat, which contains the necessary elements of a conventional spectrograph. At the output of the fibers bundle, a pseudo-slit can be formed, which is directly fed to the spectrograph. This approach offers several advantages. Firstly, it alleviates size constraints and therefore gives more flexibility to the project. In addition, few modifications to LINC need to be done, and LIINUS/SERPIL would still be able to benefit from the existing fringe tracking unit. However, there exists a crucial physical problem to investigate: fiber coupling losses.

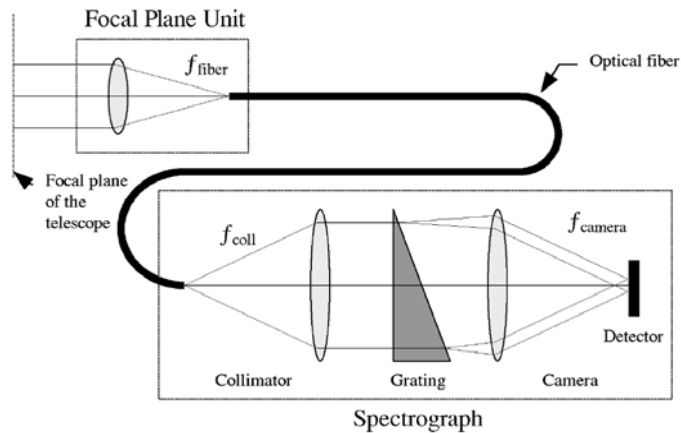

Fig. 9. Instrument concept of the Lenslet+Fibers integral-field spectrograph (adapted from [22]). In the context of the SERPIL/LIINUS project, the IFU or Focal Plane Unit is located inside the LINC cryostat

In this concept, the fibers play a very important role in the performance of the whole instrument, and therefore special attention should be given to their physical properties. The majority of fiber-fed instruments so far have been designed for operation under natural seeing conditions using multi-mode fibers [22, 23]. However, in the case of LIINUS/SERPIL, where light will be transported at ambient temperatures (see Figure 9), the thermal background at near-IR wavelengths should be as low as possible, and that can only be achieved by employing single- or few-mode fibers. The coupling to single- and few-mode fibers at the focal plane of 8m telescopes has been investigated before [24, 25]. However, this is an unexplored territory for a double-pupil configuration such as the LBT. In addition, the use of microlenses to focus the light into the fibers turns the situation to a pupil-plane coupling. Therefore, the main issue to investigate in this approach is the physics of coupling a double pupil to optical fibers.

Numerical simulations have been developed for this purpose using the Interactive Data Language (IDL) and the optical design software "Zemax". The theory, description and results of the simulations are discussed somewhere else [26]. Here we present a summary of the most important conclusions and the implications for the LIINUS/SERPIL project.

For a double-pupil configuration such as the LBT, the results of our simulations on diffraction-limited pupil-plane coupling to single- and few-mode fibers under ideal conditions can be summarized as follows:

- Single-mode fibers do not couple well the interferometric micropupil of the LBT (see the lower left panel of Fig. 10) resulting in 60% coupling efficiency

- Few-mode fibers offer higher maximum coupling efficiency (> 90%) and do not suffer much from thermal background at ambient temperatures when the number of modes is lower than 10.

- Anamorphic magnification eliminates the elongation of the interferometric electric field pattern and produces round micropupils (right panels of Fig. 10). This solution is almost equivalent to the coupling of a single telescope pupil to optical fibers.

- Two options are viable for a lenslet+fibers integral-field unit at the LBT: Anamorphic magnification together with square or hexagonal microlenses coupled to single- or few-mode fibers, and rectangular microlenses coupled to few-mode fibers with more than 3 propagating modes.

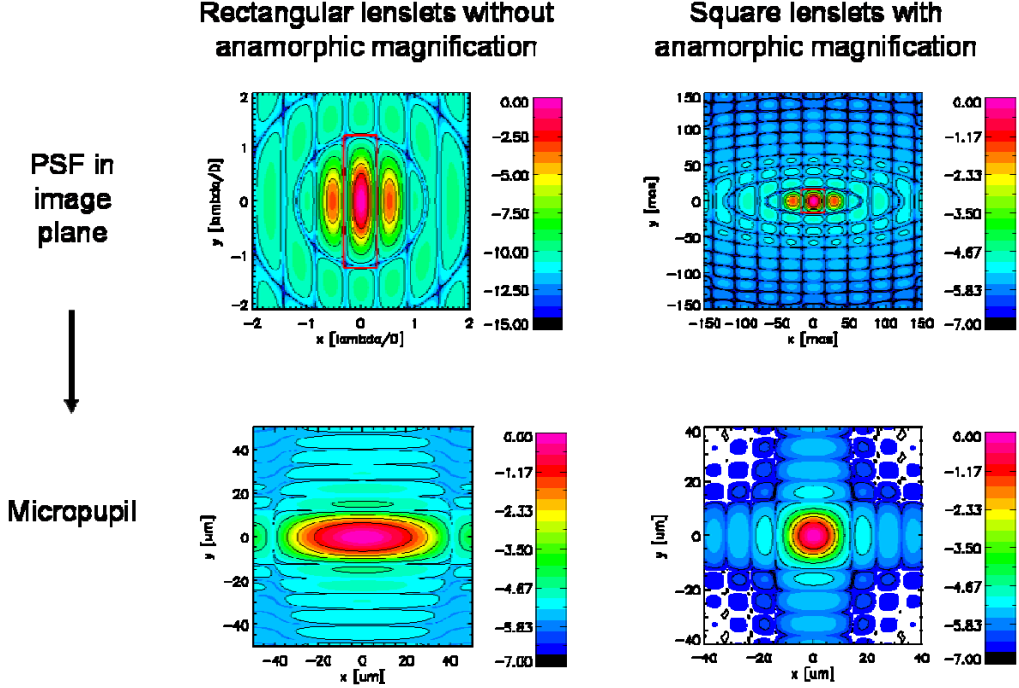

Fig. 10. Top Left: Interferometric PSF of the LBT. The rectangle delinates the central peak of the PSF and represents a possible spatial sampling using a rectangular microlens. Down Left: Micropupil obtained at the microlens focus of the filtered PSF shown above. Top Right: PSF of the anamorphic double pupil with a magnification of ~3.5 in the low resolution direction. The square delinates the central peak of the PSF and represents a possible spatial sampling using a square microlens. Down Right: Micropupil obtained at the microlens focus of the filtered PSF shown above.

As mentioned above, the coupling scenarios studied in [26] correspond to ideal cases. However, in a real astronomical observation, the coupling will be achieved most probable under non-ideal conditions. First, the incident wavefront may not be planar. Furthermore, optical power may be reflected back at the fiber's entrance (reflection loss). And principally, the fiber can be misaligned compared to the ideal position. All of these will decrease the coupling efficiency in different amounts. We have undertaken a research program to investigate the coupling under non-ideal conditions. The results are presented in subsequent publications [26, 27], which contain not only results from numerical simulations but also real measurements from a laboratory experiment.

## 5  SUMMARY AND CONCLUSIONS

We have presented developments towards the design of SERPIL/LIINUS, a near-infrared integral field spectrograph for interferometric diffraction-limited observations at the LBT. By adding such an instrument to the interferometric focus of the LBT, several science programs could be brought to a next level in terms of increased sensitivity and spatial resolution, particularly studies of stars and gas in the immediate vicinity of supermassive black holes in nearby AGN. Two instrument concepts have been studied in detail: a microlens array IFU with a spectrograph built entirely inside LINC (the LIINUS approach), and a lenslet+fibers IFU feeding an external spectrograph (the SERPIL approach). For the LIINUS concept, a preliminary design of the optical path which brings the light from LINC's original focal plane to the upper part of the cryostat, where some space is available for upgrades, has been developed. In order to fill effectively this empty volume, a very compact folded spectrograph has been designed. In the SERPIL concept, our investigations suggest that two options are viable for an instrument with a lenslet+fibers IFU: Anamorphic magnification together with square or hexagonal microlenses coupled to single- or few-mode fibers, and rectangular microlenses coupled to few-mode fibers with more than 3 propagating modes.